\documentclass[twocolumn,english,amsmath,amssymb,floatfix,superscriptaddress,longbibliography]{revtex4-2}

\usepackage[latin9]{inputenc}
\usepackage{graphicx}
\usepackage[usenames,dvipsnames]{xcolor}
\usepackage{amsmath}

\definecolor{orange}{rgb}{1,0.5,0}
\definecolor{goodgreen}{rgb}{0.1,0.5,0}
\definecolor{goodred}{rgb}{0.7,0,0}
\usepackage[colorlinks,urlcolor=goodgreen,citecolor=blue,linkcolor=goodred]{hyperref}
\usepackage{lineno}

\usepackage{tikz}
\usepackage{tocvsec2}

\newcommand{\dcom}[1]{\textcolor{NavyBlue}{\textsf{\textbf{#1}}}}

\newcommand{\comment}[1]{}

\usepackage[normalem]{ulem}

\usepackage{float}
\graphicspath{ {images/} }

\makeatother

\usepackage{babel}

\begin{document}
\preprint{\dcom{Preprint not for distribution CONFIDENTIAL | ver. of \today}}

\title{Spin-texture topology in curved circuits driven by spin-orbit interactions}

\author{A. Hijano}
\email{alberto.hijano@ehu.eus}
\affiliation{Centro de F\'isica de Materiales (CFM-MPC) Centro Mixto CSIC-UPV/EHU, E-20018 Donostia-San Sebasti\'an,  Spain}
\affiliation{Department of Condensed Matter Physics, University of the Basque Country UPV/EHU, 48080 Bilbao, Spain}

\author{E. J. Rodr\'{\i}guez}
\email{erfernandez@us.es}
\affiliation{Departamento de F\'isica Aplicada II, Universidad de Sevilla, E-41012 Sevilla, Spain}

\author{D. Bercioux}
\email{dario.bercioux@dipc.org}
\affiliation{Donostia International Physics Center (DIPC), 20018 Donostia--San Sebasti\'an, Spain}
\affiliation{IKERBASQUE, Basque Foundation for Science, Plaza Euskadi 5
48009 Bilbao, Spain}

\author{D. Frustaglia}
\email{frustaglia@us.es}
\affiliation{Departamento de F\'isica Aplicada II, Universidad de Sevilla, E-41012 Sevilla, Spain}

\begin{abstract} 
\begin{center}
    \textbf{ABSTRACT}
\end{center}
Interferometry is a powerful technique used to extract valuable information about the wave function of a system.
In this work, we study the response of spin carriers to the effective field textures developed in curved one-dimensional interferometric circuits subject to the joint action of Rashba and Dresselhaus spin-orbit interactions. By using a quantum network technique, we establish that the interplay between these two non-Abelian fields and the circuit's geometry modify the geometrical characteristics of the spinors, particularly on square circuits, leading to the localisation of the electronic wave function and the suppression of the quantum conductance. We propose a topological interpretation by classifying the corresponding spin textures in terms of winding numbers. 
\end{abstract}
\maketitle

\section*{Introduction}

Electrons subject to a cyclic motion in mesoscopic loops reveal a whole series of quantum effects, both of fundamental and practical interest. A charge circulating a magnetic flux line gathers a quantum phase leading to the  Aharonov-Bohm (AB) effect~\cite{ABeffect}, which demonstrates the distinct role played by electromagnetic potentials in quantum physics. Moreover, the AB effect is topological, meaning that it does not depend on the particulars of the loop's geometry as long as the magnetic flux line is enclosed. Aharonov-Bohm phases are nothing but an example of the geometric phases formalised by Berry~\cite{Berry_1984}, Simon~\cite{Simon_1983} and Wilczek \& Zee~\cite{Wilczek_1984} in the early 1980s, which have become increasingly influential in many areas, from condensed-matter physics and optics to high-energy and particle physics
and fluid mechanics to gravity and cosmology~\cite{Cohen_2019}. 

As for the spin degree of freedom, its coupling to magnetic and/or electric fields can lead to rich dynamics and corresponding spin quantum phases with significant consequences. One example relevant to this work is the Aharonov-Casher effect~\cite{AharonovCasher_1984}, the electromagnetic dual of the AB effect, due to the spin coupling to electric fields, \emph{i.e.}, spin-orbit interaction (SOI). Here, we focus on the combined action of Rashba and Dresselhaus SOI in polygonal and circular circuits. In semiconducting systems, the former is due to the  lack of structural inversion symmetry, whereas the latter is due to the lack of bulk inversion symmetry~\cite{Winkler_2003}.

To dig into the problem of spin dynamics in curved mesoscopic circuits, we frame the discussion in terms of field and spin textures.
By field texture, we refer to the geometry displayed by the magnetic field that couples to the spin carrier in a loop circuit, the main characteristic of which is to be inhomogeneous in direction. These fields can be external (\emph{e.g.}, inhomogeneous magnetic fields interacting through Zeeman coupling) or internal (\emph{e.g.}, effective magnetic fields emerging from SOI in curved circuits). By spin texture, we refer to the geometric shape defined by the local quantization axis of spin eigenstates along the circuit, represented in the Bloch sphere. Both field and spin textures coincide in the so-called adiabatic limit, where spin eigenstates are locally aligned with the driving field. This is the limit in which Berry geometric phases are formulated~\cite{Berry_1984}. However, reaching this limit may be either hard (due to the large fields required) or impossible (due to discontinuities in the field textures that spins cannot follow). Aharonov \& Anandan~\cite{Aharonov_1987} generalised the concept of geometric phases to the case of non-adiabatic dynamics. For $1/2$ spins, the geometric phase equals the solid angle subtended by the spin texture (times $-1/2$).  

Several works study the correlation between field and spin textures in loop circuits. It is well established that for regular (discontinuity-free) field textures, the spin dynamics is determined by the relative magnitude of two characteristic frequencies: the Larmor frequency of spin precession $\omega_{\text s}$ and the orbital frequency of carrier propagation $\omega_0$~\cite{Stern_1992,Loss_1993,Popp_2003}, where the adiabatic limit corresponds to $\omega_{\text s}/\omega_0 \gg 1$. Some works have studied how this limit is approached~\cite{Frustaglia_2001}. By assuming fully adiabatic spin dynamics in circular circuits, Lyanda-Geller demonstrated~\cite{LyandaGeller_1993} that a topological transition in a flat field texture (from a circular field texture to an oscillating one) would lead to a discontinuity in the spin Berry phase (a sudden $\pi$ shift) with observable consequences in electronic transport. Later works~\cite{Ortix_2015,Reynoso_2017} showed that this topological effect takes place far from the adiabatic limit.   

Moreover, it has been acknowledged that the geometric curvature of a circuit can play a critical role in spin dynamics~\cite{Bercioux_2005B,Ying_2016,Ying_2020,Gentile_2022,Salamone_2021}. For example, in polygonal Rashba loops where effective field-texture discontinuities at the highly curved vertices force the spin carriers to respond in a strongly non-adiabatic fashion~\cite{Bercioux_2005B,van_Veenhuizen_2006,Koga_2006,Hijano_2021,Rodriguez_2021}. This has been proven to have dramatic consequences for the correlation between field and spin textures: while field and spin textures are typically well correlated in circular Rashba loops (\emph{i.e.}, 
it usually takes a topological change in the driving field texture to produce a topological transition in the spin texture). For square Rashba loops, it has been shown that small perturbations in the field texture (created by an in-plane Zeeman field) can induce significant changes in the topological characteristics of the spin textures~\cite{Wang_2019,Frustaglia_2020,Hijano_2021}. Here, we show how similar changes can be achieved by purely electrical means | without introducing magnetic fields.  

Our study focuses on square and ring loops suitable for experimental realisation~\cite{Nagasawa_2013,Wang_2019,Nagasawa_2018}. These experiments are realised on arrays of many interferometric loops where only one single (quasi-one-dimensional) orbital mode appears to contribute to quantum interference due to the decoherence experienced by relatively slow propagating higher modes. This justifies the use of strictly one-dimensional (1D) model circuits in this work. 

In this manuscript, we study the development of spin textures and their response due to the combined action of Rashba and Dresselhaus SOI in polygonal circuits. The field textures produced by Rashba and Dresselhaus SOIs are contained within the circuit's plane. Their topology depends on the SOI components' relative magnitude, which can be controlled electrically in semiconducting nanostructures~\cite{Nitta_1997,Schapers_1998,Grundler_2000,Dettwiler_2017,Nagasawa_2018}. We find that, for specific circuit orientations, the spin textures respond with a regular pattern of topological transitions as a function of the SOIs without requiring a topological change of the driving field texture. This means that for any SOI setting, it is always possible to change the topological characteristics of the corresponding spin texture by shifting to a different setting in its vicinity. We also show how this manifests in the conductance of the circuits. 
There are several possible approaches to studying quantum transport in mesoscopic systems, such as the recursive  Green's function approach~\cite{datta1997electronic} and the tight-binding method~\cite{Groth_2014}. In this work, we address the problem by employing a quantum network (QN) technique~\cite{Kottos_1999,Gnutzmann_2006,Vidal_2000,Bercioux_2004,Bercioux_2005A,Bercioux_2005B,Ramaglia_2006}. We have recently generalised this QN technique to account for Abelian and non-Abelian gauge fields, including Rashba SOI and Zeeman fields~\cite{Hijano_2021, Rodriguez_2021}. Here, we incorporate Dresselhaus SOI and study its interplay with the Rashba SOI~\cite{Bercioux_2015}.

%
%
\begin{figure}[!t]
\centering
  \includegraphics[width=0.99\columnwidth]{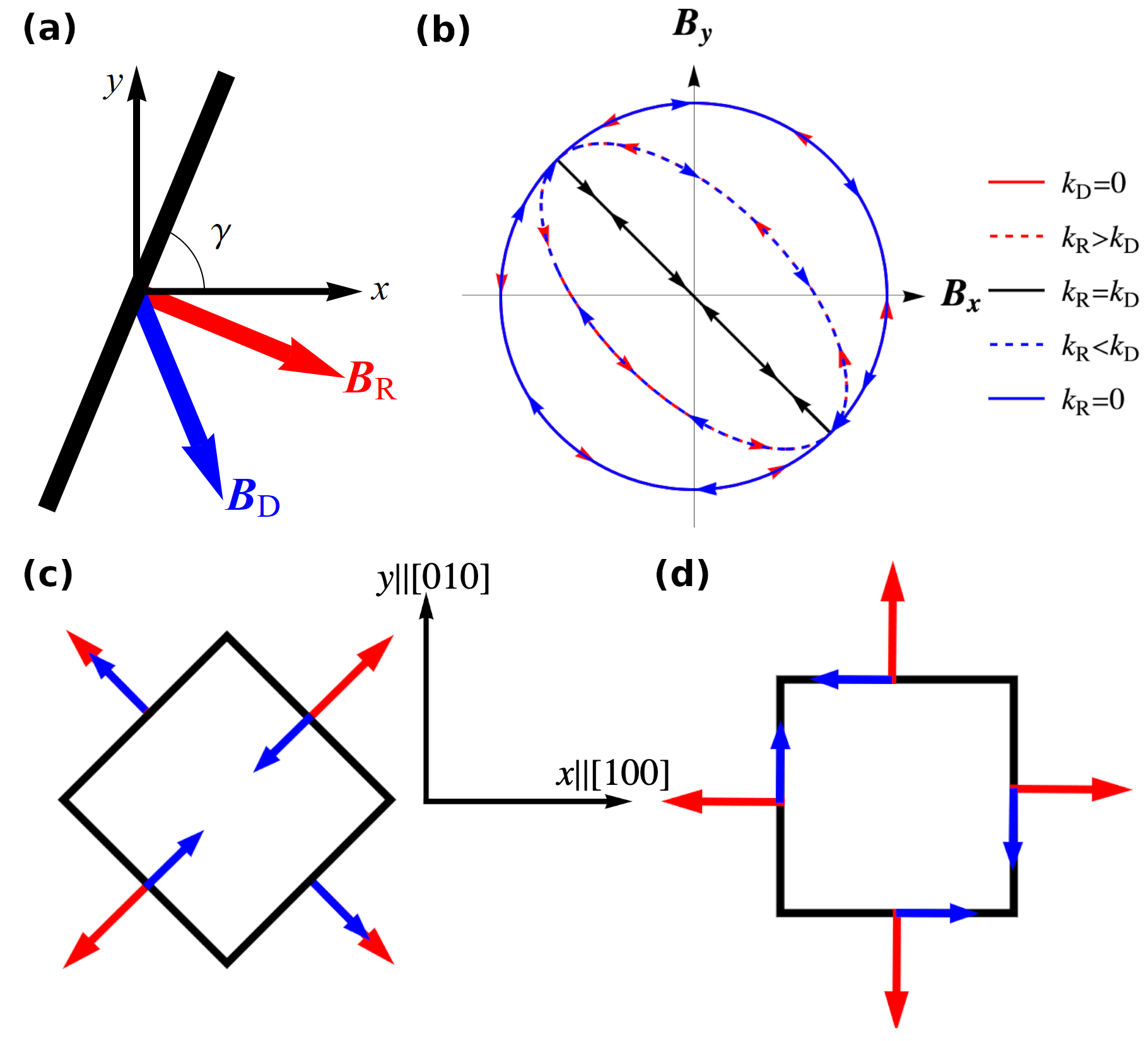}
\caption{\label{figure_1} \textbf{Effective spin-orbit interaction fields for wire, ring and square geometry:}
(a) Sketch of a quantum wire along direction $\hat{\boldsymbol{\gamma}}$ with its associated effective magnetic fields due to Rashba $\boldsymbol{B}_\text{R}$ (red arrow) and Dresselhaus $\boldsymbol{B}_\text{D}$ (blue arrow) SOIs. (b) Evolution of the effective magnetic field $\boldsymbol{B}_\mathrm{SO}=\boldsymbol{B}_\text{R}+\boldsymbol{B}_\text{D}$ for an electron moving counterclockwise around a ring for different values of the Rashba ($k_\text{R}$) and Dresselhaus ($k_\text{D}$) coupling strengths. Rashba (red) and Dresselhaus (blue) effective magnetic fields of a counterclockwise propagating spin carrier due to SOI for the case of the square circuit with (c) $\eta=0$  and (d) $\eta=\pi/4$.}
\end{figure}
%
%

\section*{Results}

\subsection*{Model and Formalism}\label{model}

We consider a 2DEG in the presence of Rashba and Dresselhaus SOIs~\cite{Rashba,Dresselhaus};  both  these terms are linear in momentum $p$~\footnote{Here, we neglect cubic corrections to Dresselhaus SOI ~\cite{Dresselhaus,Bercioux_2015}.}. Within this 2DEG, we realize single-mode 1D quantum wires along the $\hat{\boldsymbol{\gamma}}$ direction with respect to the crystallographic axes of the 2DEG | see Fig.~\ref{figure_1}(a). The quantum wire Hamiltonian then reads:
%
%
\begin{equation}\label{Hamiltonian1}
\hat{\mathcal{H}}=\frac{p^{2}}{2m^*}+\frac{\hbar k_{\mathrm{R}}}{m^*}p~(\hat{\boldsymbol{\gamma}}\times\hat{\boldsymbol{z}})\cdot\boldsymbol{\sigma}+\frac{\hbar k_{\mathrm{D}}}{m^*}p~\hat{\overline{\boldsymbol{\gamma}}}\cdot\boldsymbol{\sigma},
\end{equation}
%
%
where $k_{\mathrm{R}}$ and $k_{\mathrm{D}}$ are the Rashba and the Dresselhaus SOI strengths (in inverse-length units), respectively, $\hat{\boldsymbol{\gamma}}=(\cos\gamma,\sin\gamma,0)$ is the unit vector specifying the direction of the quantum wire, and $\hat{\overline{\boldsymbol{\gamma}}}=(\cos\gamma,-\sin\gamma,0)$ is the $y$-reflected  $\hat{\boldsymbol{\gamma}}$ (see Supplementary note 1~\cite{supplemental} for additional details). In Eq.~\eqref{Hamiltonian1}, $\boldsymbol{\sigma}$ is the vector of the Pauli matrices associated with the electron spin, $p$ is the momentum along the quantum wire,  $m^*$ is the effective electron mass of the 2DEG and $\hbar$ the reduced Planck constant.

The Rashba and Dresselhaus SOI terms in Hamiltonian~\eqref{Hamiltonian1} can be unified as
%
%
\begin{equation}\label{Hamiltonian2}
\hat{\mathcal{H}}=\frac{p^{2}}{2m^*}+\frac{\hbar \kappa}{m^*}p~ \hat{\boldsymbol{\theta}}\cdot\boldsymbol{\sigma},
\end{equation}
%
%
where
%
%
\begin{subequations}\label{kappa_theta}
\begin{align}
    \kappa&=\sqrt{k_{\mathrm{R}}^2+k_{\mathrm{D}}^2+2k_{\mathrm{R}}k_{\mathrm{D}}\sin{(2\gamma)}}, \label{kappa}\\
    \theta&=\arg\left[(k_{\mathrm{R}}\sin{\gamma}+k_{\mathrm{D}}\cos{\gamma}) \right.\nonumber \\
    & \left. \hspace{1cm}+\mathrm{i}(-k_{\mathrm{R}}\cos{\gamma}-k_{\mathrm{D}}\sin{\gamma}) \right].\label{phaseRD} 
\end{align}
\end{subequations}
%
%
Hamiltonian~\eqref{Hamiltonian2} describes an equivalent system where an electron moves along a quantum wire subject to an \emph{effective} SOI with strength $\kappa$. This SOI term can be interpreted as an effective magnetic field $\boldsymbol{B}_\mathrm{SO}=2\hbar \kappa p/(g\mu m^*)\hat{\boldsymbol{\theta}}$, where $g$ is the g-factor, $\mu$ is the Bohr magneton and $\hat{\boldsymbol{\theta}}=(\cos\theta,\sin\theta,0)$. In Fig.~\ref{figure_1}(b), we show the effective magnetic field texture experienced by spin carriers following circular trajectories for various
values of the Rashba and Dresselhaus SOIs. The arrows indicate the evolution of $\boldsymbol{B}_\mathrm{SO}$ when an electron moves counterclockwise in a circular trajectory. The field texture can be characterised topologically in terms of the winding number $\omega$ around the $z$-axis | see Eq.~\eqref{Eq_winding number}. The winding number changes depending on the relative strength between the Rashba and Dresselhaus SOIs; \emph{i.e.}, it is $\omega=1$ for $k_\mathrm{R}>k_\mathrm{D}$ and $\omega=-1$ for $k_\mathrm{R}<k_\mathrm{D}$. This change of the winding at the critical line $k_\text{R}=k_\text{D}$ is reflected in the spin texture of the polygon eigenstates, but as shown in the Section ``Topological characterization", the spin textures develop out-of-plane components that lead to richer winding patterns.
In Fig.~\ref{figure_1}(c) and~\ref{figure_1}(d) we present the SOI field texture for the two different orientations of the square circuit.
In polygonal structures, the effective field exhibits discontinuities at the vertices, but such sharp changes of direction are smoothened in realistic setups where the circuits are realized by lithographic etching of a 2DEG. If the vertices of the square are treated as slightly rounded arcs~\cite{Rodriguez_2021}, the evolution of the field texture on $\boldsymbol{B}$ space is equivalent to that of the ring [see Fig.~\ref{figure_1}(b)].

The QN problem is solved by fixing a wave function for each 1D quantum wire satisfying the Dirichlet boundary conditions. The overall solution is obtained by imposing the general boundary condition on the set of wires composing the QN | see below.
The wave function of a quantum wire can be written as~\cite{Bercioux_2004,Bercioux_2005A}:
%
%
\begin{align}\label{wavefunction1}
\boldsymbol{\Psi}(r)=\frac{\mathrm{e}^{-\mathrm{i}\kappa r \hat{\boldsymbol{\theta}}\cdot\boldsymbol{\sigma}}}{\sin(k L)}&[\sin{k(L-r)}\boldsymbol{\Psi}_\alpha \nonumber\\
&+\sin{(kr)}\mathrm{e}^{\mathrm{i}\kappa L \hat{\boldsymbol{\theta}}\cdot\boldsymbol{\sigma}}\boldsymbol{\Psi}_\beta],
\end{align}
%
%
where $\boldsymbol{\Psi}_\alpha$ and $\boldsymbol{\Psi}_\beta$ are the spinors evaluated at the quantum wire boundaries $\alpha$ and $\beta$, $k=\sqrt{2m^*\epsilon/\hbar^2+\kappa^2}$, $r$ is the coordinate along the wire, $L$ is the length of the wire and $\epsilon$ is the energy. The exponent in the prefactor of Eq.~\eqref{wavefunction1} accounts for the spin precession due to the effective magnetic field created by the two SOIs. When the Rashba and Dresselhaus SOI strengths are equal, and $\gamma=3\pi/4$, both SOI terms cancel in the Hamiltonian, the effective magnetic field vanishes, and the energy dispersion turns spin degenerate~\cite{Schliemann_2003A,Schliemann_2003B}. In this case, $\kappa$ vanishes, so the SU(2) rotation disappears from the wave function. This means that the spinor evolves along the wire as a free particle.

The spin-carrier dynamics in a QN can be solved by considering two general boundary conditions: first, the continuity of the wave function to each node of the QN, and second by requiring the conservation of the probability current at the same points~\cite{Gnutzmann_2006}.\\
Equation~\eqref{wavefunction1} together with the conservation of the current at the vertices provides the values of the wave function at the vertices $\boldsymbol{\Psi}_{\alpha}$, and $\boldsymbol{\Psi}(r)$ by extension. \\ We can evaluate the transport properties by supplementing the QN with an additional connection to external leads. The extension of the method is explained in the Methods section.

%
%
\begin{figure*}[!ht]
    \centering
    \includegraphics[width=0.89\textwidth]{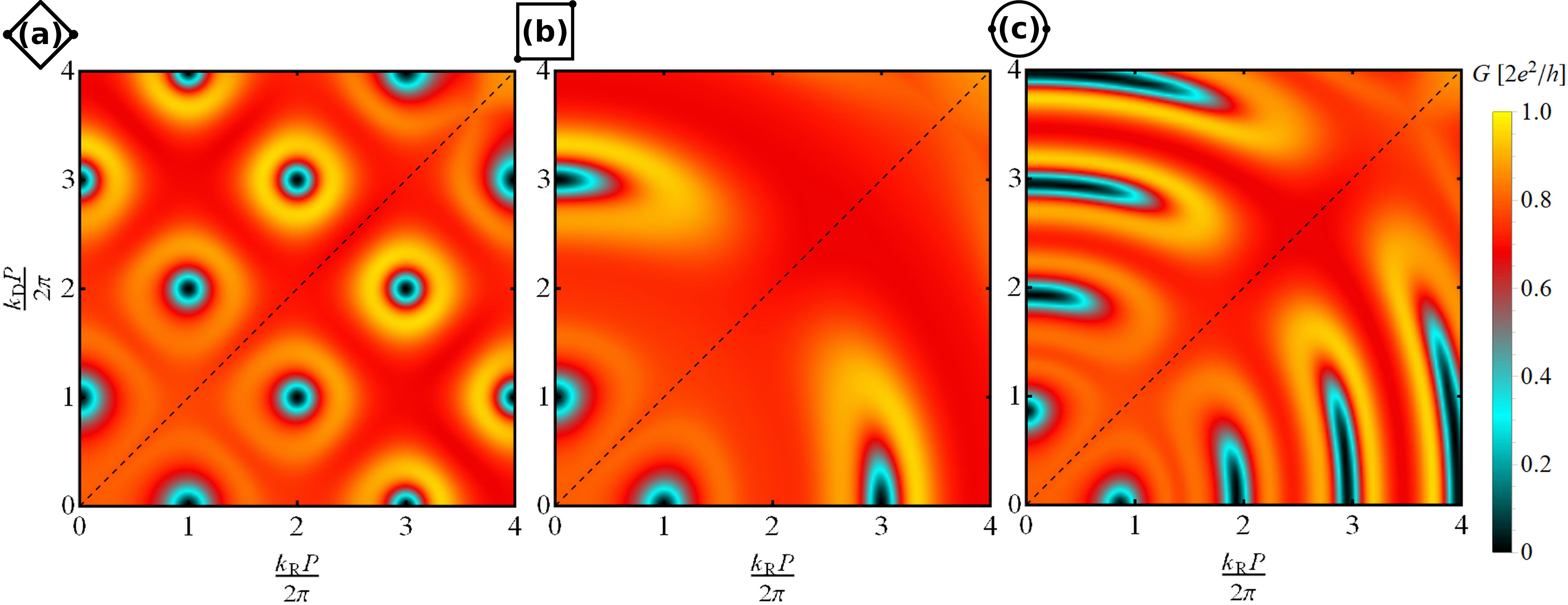}
    \caption{\label{conductance1}\textbf{Conductance for the square and the ring geometries: } Conductance $G$ in units of $2e^2/h$ as a function of the dimensionless Rashba and Dresselhaus spin-orbit interaction for a square for different orientations (a) $\eta=0$, (b) $\eta=\pi/4$ and a (c) ring with perimeter $P$. In all the panels, the dashed line corresponds to the critical line, where $k_\text{R}=k_\text{D}$. The left (right) dot in the sketches of the polygonal structures indicates the position of the input (output) lead.}
\end{figure*}
%
%

\subsection*{Conductance pattern}\label{conductance}

In this section, we study the transport properties of square and ring loops by applying the QN formalism. Rings are modelled as regular polygons of perimeter $P$ with a large number of edges ($N \gg 1$)~\cite{Bercioux_2005B} such that the Fermi wavelength and the spin precession length of the carriers are much larger than the edges' length $L=P/N$. This overcomes the problem of dealing with approximate solutions for rings subject to Rashba and Dresselhaus SOI \cite{Lia_2021, Lia_2022}. These restrictions do not apply to square loops. Moreover, mesoscopic experiments are typically carried out in the so-called semiclassical regime where the Fermi wavelength of the carriers is much smaller than the loops' perimeter, such that $k \gg 2\pi/P$ \cite{Nagasawa_2012,Nagasawa_2013,Wang_2019,Frustaglia_2020}. We calculate the ballistic conductance of these polygonal structures by using the Landauer approach~\cite{landauer_formula,supplemental} [see Eq.~\eqref{landauer_conductance}].

In Fig.~\ref{conductance1}, we show the conductance maps as a function of the dimensionless SOI intensities
$k_\text{R}P/2\pi$ and $k_\text{D}P/2\pi$ for different interferometric loops.  Figures~\ref{conductance1}(a) and~\ref{conductance1}(b) show the conductance for the case of square-shaped loops with different orientations, with $\eta$ the rotation angle measured from the ``diamond" configuration depicted in Fig.~\ref{conductance1}(a) (corresponding to square's sides forming an angle of $\pi/4$ with respect to the crystallographic axes). In Fig.~\ref{conductance1}(c), instead, we show the conductance corresponding to a ring-shaped loop. For all configurations, the conductance shows a symmetric behaviour with respect to the critical line $k_\text{R}=k_\text{D}$ along which the Rashba and Dresselhaus SOIs have the same strength.
This can be understood in terms of the Hamiltonian in Eq.~\eqref{Hamiltonian2}. If the strengths of the interactions are interchanged, $k_{\mathrm{R}} \leftrightarrow k_{\mathrm{D}}$, the value of the equivalent effective SOI strength $\kappa$ for a given edge remains unchanged, but the direction of the effective magnetic field $\boldsymbol{B}_\mathrm{SO}$ becomes $\theta'=3\pi/2-\theta$. This means that the $\boldsymbol{B}_\mathrm{SO}$ field texture of the polygon is mirrored with respect to the same $\hat{r}_-=(\hat{x}-\hat{y})/\sqrt{2}$ direction for all edges, so the conductance pattern remains unchanged under the $k_{\mathrm{R}} \leftrightarrow k_{\mathrm{D}}$ transformation.

The most interesting case is presented in the case of Fig.~\ref{conductance1}(a): the conductance presents a checkerboard pattern. In the absence of Dresselhaus SOI, the minima of the conductance are presented every $k_\text{R}P/2\pi=2n+1$ with $n\in\mathbb{N}_0$~\cite{Bercioux_2005B,Hijano_2021,Rodriguez_2021}. A similar behaviour is observed in the absence of Rashba SOI with $k_\text{R}$ replaced by~$k_\text{D}$. The combined presence of the two SOIs adds an overall shift of the conductance minima by a factor of $2\pi$. The resulting checkerboard pattern was first reported in Ref.~\cite{Ramaglia_2006}. In the next section, we elaborate on this by studying the geometric properties of spinors in terms of winding numbers. 

By rotating the square with respect to the crystallographic axes | Fig.~\ref{conductance1}(b) | we find something remarkable: the checkerboard conductance pattern disappears. Instead, we find a conductance map that looks similar to that of a ring-shaped loop | Fig.~\ref{conductance1}(c) | except for a period-doubling due to strongly non-adiabatic processes at the square vertices hindering spin-phase development ~\cite{Frustaglia_2004,Bercioux_2005B}.

\begin{figure}[!t]
    \includegraphics[width=0.9\columnwidth]{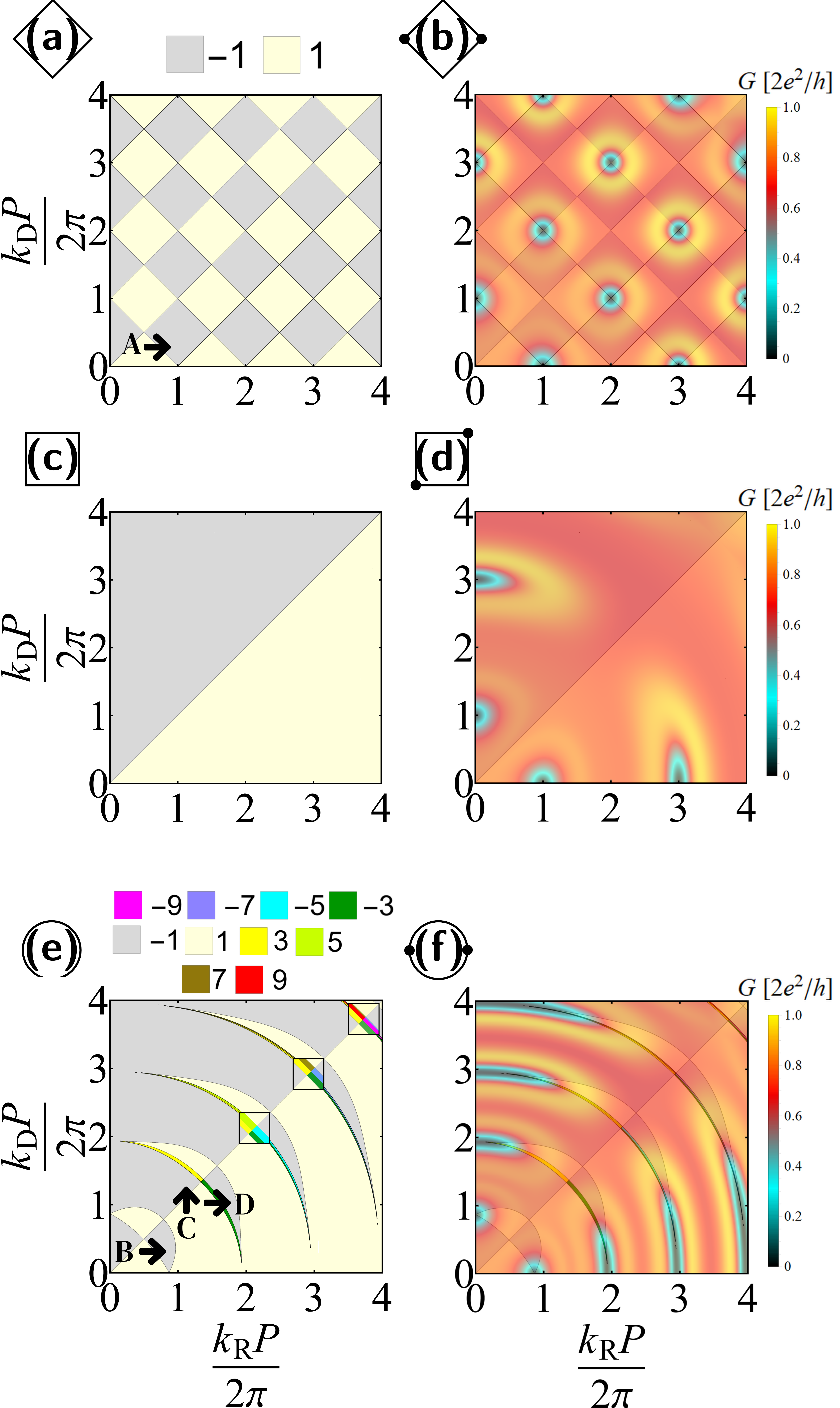}
    \caption{\label{Winding1} \textbf{Comparison of the winding number and conductance for the square and the ring geometries: } Winding number $\omega$ (left column) and winding number overlapped with the conductance in units of $2e^2/h$ (right column) for a square (a-b), a square rotated $\pi/4$ from the initial configuration (c-d), and  a ring (e-f) subject to Rashba and Dresselhaus spin-orbit interaction. The left (right) dot in the sketches of the polygonal structures in panels (b), (d), and (f), indicates the position of the input (output) lead. The insets in (e) represent zoom in the anomalous winding in the critical line for better visualization of its structure.}
\end{figure}

From Fig.~\ref{figure_1}(c) and~\ref{figure_1}(d), we observe that the SOI field textures present discontinuities at the vertices of the squares. For $\eta=0$, Fig.~\ref{figure_1}(c), the discontinuities have an angle $\pi/2$ for any relative field strength (except when Rashba and Dresselhaus SOIs are equal and a persistent spin helix is set up). As $\eta$ increases, field discontinuities are softened, spin scattering is discouraged, and destructive spin interference is suppressed: optimal $\pi/2$ field-texture discontinuities at the vertices persist only along the Rashba and Dresselhaus axes in Fig.~\ref{conductance1}(b), where a definite interference pattern in the conductance survives as discussed in Refs.~\cite{Bercioux_2005A,Hijano_2021,Rodriguez_2021} for the Rashba case. 
In the Supplementary note 2~\cite{supplemental}, we show how the conductance and the winding number of the propagating spin modes evolve from the checkerboard pattern in Fig.~\ref{Winding1}(a) and~\ref{Winding1}(b) to the simpler structure in Fig.~\ref{Winding1}(c) and~\ref{Winding1}(d) upon changing the orientation of the square.
 \\
For the case of a ring, Fig.~\ref{conductance1}(c), the conductance presents a fishbone structure with minima as a function of the Rashba SOI following the sequence predicted theoretically in Ref.~\cite{Frustaglia_2004}. Similar behaviour is observed as a function of the Dresselhaus SOI term. However, in the presence of both SOI terms, the conductance behaviour is more intricate.

Interestingly, the conductance remains constant along the critical line regardless of the orientation or number of edges of the polygon [see Supplementary note 3~\cite{supplemental} for the additional case of a hexagonal and octagonal loop]. When the Rashba and Dresselhaus SOIs have the same strength, the effective SOI field decouples from the momentum and points always in the same direction~\cite{Schliemann_2003A,Schliemann_2003B}. Moreover, the spin precession angle only depends on the distance travelled along the $\hat{r}_+=(\hat{x}+\hat{y})/\sqrt{2}$ direction. 
This effect is known as \emph{persistent spin helix}~\cite{Bernevig}, and in recent years several experiments claimed to have achieved this effect~\cite{Koralek_2009,Kohda_2012,Walser_2012,Kunihashi_2016,Dettwiler_2017}.
Since all paths contributing to the transmission amplitude begin and end at the same points, the spin precession along each path is the same, and the resulting interference is always constructive.

%
%
%
%

\subsection*{Topological characterization} \label{Winding number}

We can characterise spin and field textures topologically in terms of (integer) winding numbers around the $z$-axis. This quantity is defined as
%
%
\begin{align}\label{Eq_winding number}
\omega=\frac{1}{2 \pi} \int_{0}^{P} d \ell\left(\widehat{\boldsymbol{n}} \times \frac{d \hat{\boldsymbol{n}}}{d \ell}\right) \cdot \hat{\boldsymbol{z}}
\end{align}
%
%
with $\hat{\boldsymbol{n}}(\ell)$ a in-plane unit vector and $0\le \ell\le P$ a linear parametrization of the circuit's perimeter. For the field texture, we identify $\hat{\boldsymbol{n}}(\ell)$ with $\hat{\boldsymbol{\theta}}(\ell)$ in Eq.~\eqref{Hamiltonian2}. This means that $\omega=1$ for dominating Rashba SOI and $\omega=-1$ for dominating Dresselhaus SOI [see Fig.~\ref{figure_1}(b)]. The transition occurs at the critical line $k_{\text R}=k_{\text D}$. As for the spin texture, this is given by $\hat{s}(\ell)=\langle \Psi(\ell)|\boldsymbol{\sigma}|\Psi(\ell)\rangle$. In this case, $\hat{\boldsymbol{n}}$ is identified with the normalized projection of $\hat{s}(\ell)$ on the $xy$-plane. 

Spin textures developed in Rashba and Dresselhaus squares present a periodic, checkerboard-like pattern alternating positive and negative windings | see Fig.~\ref{Winding1}(a). This pattern contrasts with the simplicity of the field texture driving the spin dynamics discussed in Section ``Model and Formalism", demonstrating the possibility of producing topological transitions in the spin texture by slightly tuning the SOI fields. This means that the spin winding can change from clockwise (CW) to counterclockwise (CCW) and vice versa without changing the winding of the field, except for the particular orientation of $\eta=\pi/4$, Fig.~\ref{Winding1}(c), where the field and spin textures stay fully correlated. Figure~\ref{Winding1}(b) shows that the winding pattern is fully correlated with the conductance up to a period-doubling.

%
%
\begin{figure}[!t]
\centering
    \includegraphics[width=0.75\columnwidth]{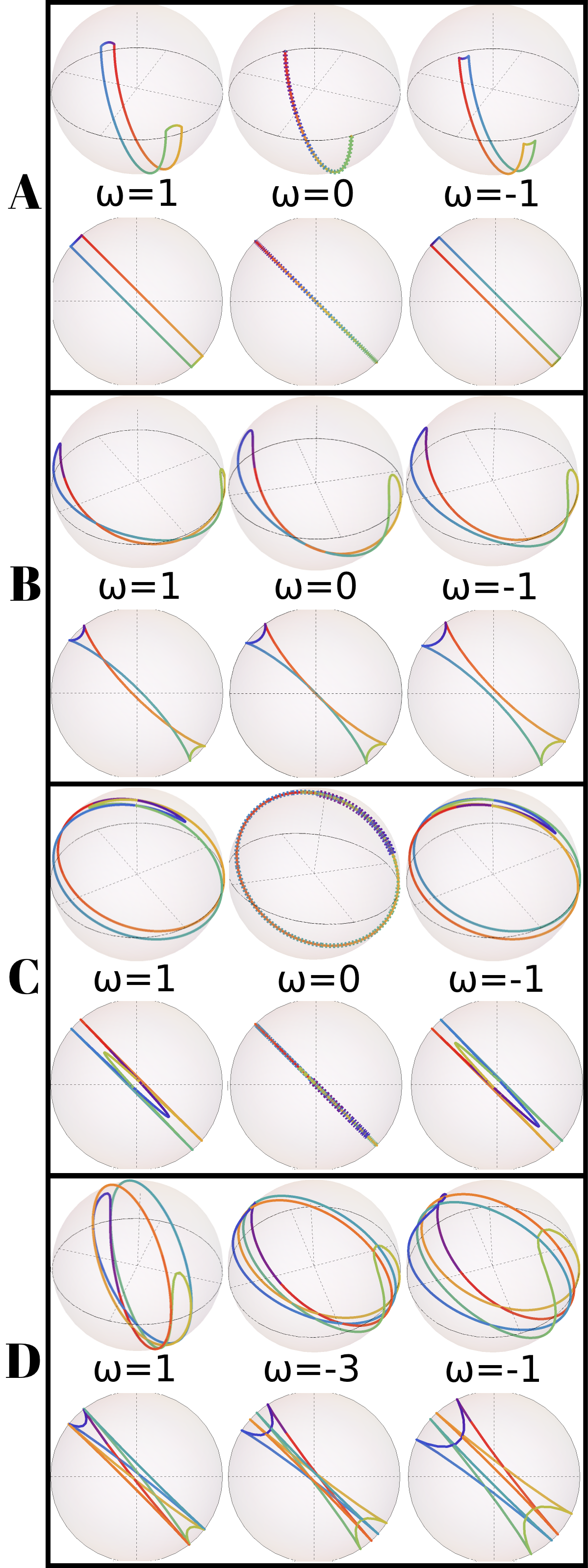}
    \caption{\label{Textures}\textbf{Spin texture for the square and the ring geometries: } The spin texture of a propagating mode in the Bloch sphere (up) and its azimuthal projection (down) for different spin-orbit interaction strengths. The texture is evaluated for three values of the winding number ($\omega$), covered by the arrows A-D in Fig.~\ref{Winding1}, with corresponding strengths of SOI taken from the back of the arrow to the tip. In each panel, the color indicates the circulation of the local spin states as the carrier propagates along the perimeter from red to violet.}
\end{figure}
%
%

In Fig.~\ref{Textures}(A), we present a series of spin textures undergoing a topological transition along segment A in Fig.~\ref{Winding1}(a), corresponding to a square loop. We find that a positive winding texture becomes negative by collapsing into a flat texture (subtending no solid angle and no geometric phase) at the critical line~\cite{Rodriguez_2021}.

In the case of ring loops, Fig.~\ref{Winding1}(e) and \ref{Winding1}(f), we find that the spin textures have a dominant tendency to follow the driving field texture by sharing its topological characteristics. Still, we find a fishbone pattern of anomalous winding | see insets in Fig.~\ref{Winding1}(e), where we observe that the winding gets values larger than $\pm1$. To shed some light on it, it is useful to assume a dominating Rashba SOI with $k_{\text R} \gg k_{\text D}$ and work in the rotating frame where the radial Rashba texture is uniform (with an oscillating Dresselhaus perturbation). We notice that the fishbone pattern meets the Rashba axis at points $k_{\text D}P=\pi\sqrt{4j^2-1}$ with $j$ integer. This coincides with the Rabi condition for spin resonance in the rotating frame. As the magnitude of Dresselhaus perturbation increases, the resonance condition changes by undergoing a so-called Bloch-Siegert shift~\cite{Bloch_1940,Reynoso_2017}. Something similar happens along the Dresselhaus axis. Close to the resonance condition, complex spin textures emerge with anomalous winding. Figures~\ref{Textures}(B),~\ref{Textures}(C), and~\ref{Textures}(D) illustrate the winding transitions taking place in these textures. We notice that, in contrast to the case of square loops, a winding transition does not require a full collapse of the spin texture with vanishing geometric phases. 
Still, in both square and ring geometries, the spin winding is antisymmetric with respect to the critical line along which the driving field changes topology.

\section*{Conclusions and Outlook}\label{ConclusionsandOutlook}

We demonstrate how the geometry of SOI circuits can be used to manipulate the carriers' spin state. Effective SOI field textures are built by introducing circuit sections of different curvatures steering the carriers' momentum. In this way, highly curved vertices in polygon circuits act as effective spin-scattering centers for the carriers. This can be achieved by purely electrical means (without introducing magnetic fields that break time-reversal symmetry), in contrast to other proposals \cite{Wang_2019,Hijano_2021}.  

For square circuits subject to Rashba and Dresselhaus SOI, we find that the topological characteristics of the spin textures can be manipulated with relative ease by electric control of the SOIs in semiconducting nanostructures  \cite{Nitta_1997,Schapers_1998,Grundler_2000,Dettwiler_2017,Nagasawa_2018}. This contrasts with the case of ring circuits where, as a general rule, a topological change in the field texture is required to induce a corresponding change in the spin textures. Still, this restraint can be overcome by tuning the SOI to satisfy the spin resonance conditions where complex spin textures develop. Moreover, additional possibilities for spin control appear by in-plane rotation of square circuits with respect to the crystallographic axes. 

Remarkably, these topological features leave an imprint on the quantum conductance of the circuits, which can be addressed experimentally. We find a correlation between the spin-dependent conductance and a winding number associated with the propagating spin modes. This demonstrates that conventional conductance measurements can reveal the geometrical properties of the spin-carrier states.

\section*{acknowledgments}

D.B. acknowledges the support from the Spanish MICINN-AEI through Project No.~PID2020-120614GB-I00 (ENACT), the Transnational Common Laboratory $Quantum-ChemPhys$, 
the funding from the Basque Government's IKUR initiative on Quantum technologies (Department of Education), and from the Gipuzkoa Provincial Council within the QUAN-000021-01 project. A.H. acknowledges funding from the University of the Basque Country (Project~PIF20/05) and financial support from Spanish MICINN-AEI through project No.~PID2020-114252GB-I00 (SPIRIT) and the Basque Government (grant No.~IT-1591-22). D.F. and E.J.R. acknowledge support from the Spanish MICINN-AEI through Project No.~PID2021-127250NB-I00 (e-QSG) and from the Andalusian Government through PAIDI 2020 Project No. P20-00548 and FEDER Project No.~US-1380932. We thank A.A. Reynoso for valuable comments regarding winding-number computation.  

\section*{Methods}

\subsection*{Formalism for quantum transport}\label{transportformalism}

\begin{figure}[!t]
    \centering
    \includegraphics[width=0.9\columnwidth]{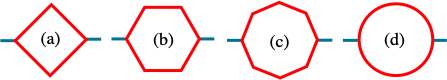}
    \caption{\label{polygons} \textbf{Various intereferometric geometries: } Sketch of the polygonal structures considered  for the quantum transport: (a) square, (b) hexagon, (c) octagon, (d) ring.}
\end{figure}

Here we present the quantum network (QN) formalism used to study the transport properties of polygonal QNs, Fig.~\ref{polygons}. Semi-infinite input and output leads are attached to the network's vertices for the transport measurements. Each lead consists of a quantum wire with two spin channels. The leads are not subjected to any interaction, so they are characterised at zero temperature by the Fermi energy and a wave vector $k$. We assume that the channels behave like incoherent sources, so there is no phase relationship between the different input channels~\cite{landauer_formulation}.

In a system with $N_{\mathrm{in}}$ ($N_{\mathrm{out}}$) input (output) channels, if an electron is injected through input channel $\sigma$ with wavenumber $k$, the wave function alongside the channels can be written as
%
%
\begin{subequations}
\begin{equation}\label{input}
\Psi_{\mathrm{in},\sigma'}(r)=\mathrm{e}^{\mathrm{i}k r}\delta_{\sigma'\sigma}+r_{\sigma'\sigma}\mathrm{e}^{-\mathrm{i}k r},
\end{equation}
%
%
\begin{equation}\label{output}
\Psi_{\mathrm{out},\sigma'}(r)=t_{\sigma'\sigma}\mathrm{e}^{\mathrm{i}k r},
\end{equation}
\end{subequations}
%
%
where $r$ is the position measured from the edge, and it is negative for input leads and positive for output leads. Here $r_{\sigma'\sigma}$ and $t_{\sigma'\sigma}$ are the channel-resolved reflection and transmission coefficients, respectively, so that $\sum_{\sigma'}^{N_{\mathrm{in}}}\left|r_{\sigma'\sigma}\right|^2+\sum_{\sigma'}^{N_{\mathrm{out}}}\left|t_{\sigma'\sigma}\right|^2=1$. The indices $\sigma$ and $\sigma'$ specify the lead and the spin state of the channel. We define the total transmission and reflection coefficients of a channel $\sigma$ as
%
%
%
%
%
%
\begin{equation}
    T_{\sigma}=\sum_{\sigma'}\left|t_{\sigma\sigma'}\right|^2\; , \qquad R_{\sigma}=\sum_{\sigma'}\left|r_{\sigma\sigma'}\right|^2\; .
\end{equation}
%
%
where the sum runs over the input channels. The total transmission (reflection) is given by the sum of the transmission (reflection) coefficients of the output (input) channels,
%
%
\begin{subequations}
    \begin{align}
        T&=\sum_{\sigma}T_{\sigma}=\sum_{\sigma\sigma'}\left|t_{\sigma\sigma'}\right|^2\\
        R&=\sum_{\sigma}R_{\sigma}=\sum_{\sigma\sigma'}\left|r_{\sigma\sigma'}\right|^2
    \end{align}
\end{subequations}
%
%

The zero-temperature conductance $G$ based on the Landauer formula reads~\cite{datta1997electronic}:
%
%
\begin{equation}\label{landauer_conductance}
    G=\frac{e^2}{h}\mathrm{Tr}\: \left[tt^{\dagger}\right]=\frac{e^2}{h} T.
\end{equation}
%
%
Equation~\eqref{landauer_conductance} sets an upper limit for the conductance, which is bounded by the number of input channels, such that $G \leq N_{\mathrm{in}} e^2/h$.

The wavefunction of the quantum network satisfies boundary conditions at the vertices, which ensure the continuity (uniqueness) of the wavefunction and the conservation of the probability current. In an isolated quantum network, imposing the continuity of the wave function and conserving the probability current yields a set of linear homogeneous equations where the variables are the values of the wave function at the vertices. This allows us to study the spectral properties of the quantum network. When adding the external leads, the system's energy is fixed by the Fermi energy of the leads. Due to the first term in Eq.~\eqref{input}, the set of equations becomes inhomogeneous, with a unique solution for $T$ and $R$.

In a system with Rashba and Dresselhaus spin-orbit interactions (SOIs), the wave function of a wire is described by the values it takes at the nodes $\boldsymbol{\Psi}_\alpha$ [see Eq.~\eqref{wavefunction1}]. The single-valuedness of the wave function at the nodes is automatically satisfied by this equation. In addition, imposing the continuity of the wave function at the vertices connected to external leads allows writing the reflection and transmission coefficients of the leads in terms of $\boldsymbol{\Psi}_\alpha$. Therefore, the number of unknown variables equals the number of vertices $V$. The conservation of probability current at the nodes allows one to write $V$ equations, which fix the values of $\boldsymbol{\Psi}_\alpha$, and consequently the reflection/transmission coefficients. Notice that the presence of SOI modifies the definition of probability current~\cite{Hodge_2014}. This is accounted for by the extended derivative:
%
%
\begin{equation}
D=\frac{\partial}{\partial r}+\mathrm{i}k_{\mathrm{R}}(\hat{\boldsymbol{\gamma}}\times\hat{\boldsymbol{z}})\cdot\boldsymbol{\sigma}+\mathrm{i}k_{\mathrm{D}}\hat{\overline{\boldsymbol{\gamma}}}\cdot\boldsymbol{\sigma}=\frac{\partial}{\partial r}+\mathrm{i}\kappa\hat{\boldsymbol{\theta}}\cdot\boldsymbol{\sigma}.
\end{equation}
%
%
The conservation of probability current at a node is given by the sum of the outgoing extended derivatives of the wave function, which must be equal to zero. For a generic node $\alpha$, the continuity of probability current reads
%
%
\begin{equation}\label{probability current}
\sum_{\langle\alpha,\beta\rangle} \left. D\boldsymbol{\Psi}_{\alpha,\beta}(r) \right|_{r=0}=0\; ,
\end{equation}
%
%
where the sum $\sum_{\langle\alpha,\beta\rangle}$ runs over all nodes $\beta$ which are connected to $\alpha$. This equation can be rewritten in terms of $\boldsymbol{\Psi}_{\alpha}$ and $\boldsymbol{\Psi}_{\beta}$. For internal nodes, it reads
%
%
\begin{equation}\label{current conservation 1}
\boldsymbol{M}_{\alpha,\alpha}\boldsymbol{\Psi}_{\alpha}+\sum_{\langle\alpha,\beta\rangle}\boldsymbol{M}_{\alpha,\beta}\boldsymbol{\Psi}_{\beta}=0\; ,
\end{equation}
%
%
where
%
%
\begin{subequations}
\begin{align}
    \boldsymbol{M}_{\alpha,\alpha}&=\sum_{\langle\alpha,\beta\rangle}\frac{k_{\beta,\alpha}}{\tan{k_{\beta,\alpha}L}}\\
    \boldsymbol{M}_{\alpha,\beta}&=-\frac{k_{\beta,\alpha}}{\sin{k_{\beta,\alpha}L}}\mathrm{e}^{\mathrm{i}\kappa_{\beta,\alpha} L \hat{\boldsymbol{\theta}}_{\beta,\alpha}\cdot\boldsymbol{\sigma}}\; .
\end{align}
\end{subequations}
%
%
Here $\kappa_{\beta,\alpha}$ and $\hat{\boldsymbol{\theta}}_{\beta,\alpha}$ indicate the strength and direction of $\boldsymbol{B}_\mathrm{SO}$ for an electron travelling from vertex $\alpha$ towards vertex $\beta$, see Eq.~\eqref{kappa_theta}.

\subsection*{Computation of the winding number}\label{Windingmodel}
The model used to compute the winding number was built upon the one used in \cite{Rodriguez_2021} where we consider a regular polygon of $N$ conducting sides of length $L = P/N$ with $P$ being the perimeter, which lies on the $xy$-plane. Each side connects the vertices $\alpha$ and $\beta$ and it is oriented along directions $\hat{\gamma}_{\beta,\alpha}$ (from $\alpha$ to $\beta$). The spin-carrier dynamics along each side are determined by Hamiltonian~\eqref{Hamiltonian2}. The SOI terms can be interpreted as an effective in-plane magnetic field $\boldsymbol{B}_\mathrm{SO}=2\hbar \kappa p/(g\mu m^*)\hat{\boldsymbol{\theta}}$ [see Eq.~\eqref{kappa_theta}] coupled to the itinerant spins. \\
%
%
\begin{figure}[!ht]
\centering
 \includegraphics[width=0.75\columnwidth]{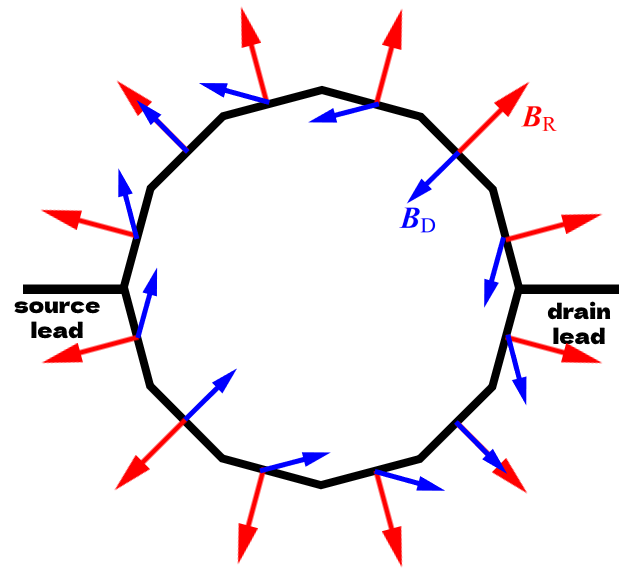}
\caption{\textbf{Rashba and Dresselhaus spin-orbit interaction fields on a  dodecagon geometry: } One-dimensional polygon model with its associated effective magnetic fields due to Rashba $\boldsymbol{B}_\text{R}$ (red arrows) and Dresselhaus $\boldsymbol{B}_\text{D}$ (blue arrows) SOIs. It shows the effective fields corresponding to counterclockwise propagating spin carriers.}\label{phaseSOI}
\end{figure}
%
%
The solutions of the Schr\"{o}dinger equation are plane waves propagating along each side from vertex $\alpha$ towards vertex $\beta$ as 
%
%
\begin{equation}
    |\psi(r)\rangle=\mathrm{e}^{-\mathrm{i} k_\text{F} r} \mathrm{e}^{-\mathrm{i} \kappa r\hat{\boldsymbol{\theta}}_{\beta,\alpha} \cdot \boldsymbol{\sigma}}|\psi(0)\rangle,
\end{equation}
%
%
with $k_\text{F}$ the Fermi wavenumber. The first prefactor corresponds to the kinetic phase of the carrier associated with the dynamics of charged particles, while the second prefactor represents the spin phase due to spin precession. The propagation of a spin carrier from $\alpha$ to $\beta$ is then fully determined by the phases $ k_\text{F} L+\kappa L\hat{\boldsymbol{\theta}}_{\beta,\alpha} \cdot \boldsymbol{\sigma}$, and the spin evolution along one side is determined by the momentum-independent spin rotation operator: 
%
%
\begin{equation}
    R_{\beta,\alpha}=\exp \left[-\mathrm{i} \kappa L\hat{\boldsymbol{\theta}}_{\beta,\alpha} \cdot \boldsymbol{\sigma}\right] 
\end{equation}
%
%
with $R_{\beta,\alpha}^\dagger=R_{\alpha,\beta}$ due to time-reversal symmetry.

The full spin evolution along counterclockwise (CCW) and clockwise (CW) propagating paths from vertex 1 to vertex $N$ is then given by the unitary operators 
%
%
\begin{equation}
    U_{+}(N)=R_{1 N} \ldots R_{32} R_{21}
\end{equation} 
%
%
and
%
%
\begin{equation}
   U_{-}(N)=R_{12} \ldots R_{N-1, N} R_{N 1}
\end{equation}
%
%
with $U_{-}(N)=U_{+}^\dagger (N)$, see Fig~\ref{phaseSOI}. 

The spin rotation operator allows us to obtain the $xy$-projection of the spin texture as $\hat{\mathbf{s}}_{x,y}(r)=\left\langle\chi_{s}|\boldsymbol{\sigma_{x,y}}| \chi_{s}\right\rangle$, then it is possible to compute the angle accumulated around the $z$-axis by the itinerant spin state as the carrier propagates along each segment as the phase of the complex number $s_{x}(r)+\mathrm{i}s_{y}(r)$.

Completing a CCW round trip, we obtain the accumulated angle around the $z$-axis and, therefore the winding number.

\section*{Data availability}
Numerical data used to generate all the figures in this manuscript is available upon reasonable request.

\section*{Code availability}
The codes employed in this study are available from the authors on reasonable request.

\section*{Author Contributions}
A.H. and E.J.R. contributed equally to this work. A.H., E.J.R., D.B. and D.F. contributed to writing the manuscript.

\section*{Competing interests}
The authors declare no competing interests. Dario Bercioux is an Editorial Board Member for Communications Physics, but was not involved in the editorial review of, or the decision to publish this article.


%

\cleardoublepage
\numberwithin{equation}{section}
\setcounter{figure}{0}
\global\long\def\theequation{S\arabic{section}.\arabic{equation}}
\global\long\def\thefigure{S\arabic{figure}}
\global\long\def\thesection{Supplementary note \arabic{section}}
\global\long\def\thesubsection{\Alph{subsection}}

\begin{widetext}
\begin{center}
\textbf{\large{}Supplemental material for ``Spin-texture topology in curved circuits driven by spin-orbit interactions''}{\large{} }
\par\end{center}{\large \par}

\end{widetext}

\section{Dresselhaus SOI terms in the Hamiltonian}
Here we show that Rashba and Dresselhaus terms of Eq.~(1) in the main text are equivalent to the standard Hamiltonian present in the literature~\cite{Ganichev_2003,Schliemann_2017}. In the main text, we set the coordinates along cubic axes, $x\parallel[100]$, $y\parallel[010]$ and $z\parallel[001]$\\
%
%
\begin{subequations}
\begin{align}
H_{\mathrm{SO}}&=\frac{\hbar k_{\mathrm{R}}}{m^*}p~(\hat{\boldsymbol{\gamma}}\times\hat{\boldsymbol{z}})\cdot\boldsymbol{\sigma}+\frac{\hbar k_{\mathrm{D}}}{m^*}p~\hat{\overline{\boldsymbol{\gamma}}}\cdot\boldsymbol{\sigma} \nonumber \\
&=\frac{\hbar k_{\mathrm{R}}}{m^*}~(p_y\sigma_x-p_x\sigma_y)+\frac{\hbar k_{\mathrm{D}}}{m^*}~(p_x\sigma_x-p_y\sigma_y) \label{EqS1A}\\
&=\frac{\hbar k_{\mathrm{R}}}{m^*}~(p_{y'}\sigma_{x'}-p_{x'}\sigma_{y'})+\frac{\hbar k_{\mathrm{D}}}{m^*}~(p_{x'}\sigma_{y'}+p_{y'}\sigma_{x'})\; \label{EqS1B},
\end{align}
\end{subequations}
%
%
where in the Eq.~\eqref{EqS1B}, we have made a coordinate transformation to the coordinate system $x'\parallel[1\bar{1}0]$, $y'\parallel[110]$ and $z'\parallel[001]$.

\section{Case of the square}
%
%
\begin{figure}[!h]
    \includegraphics[width=0.8\columnwidth]{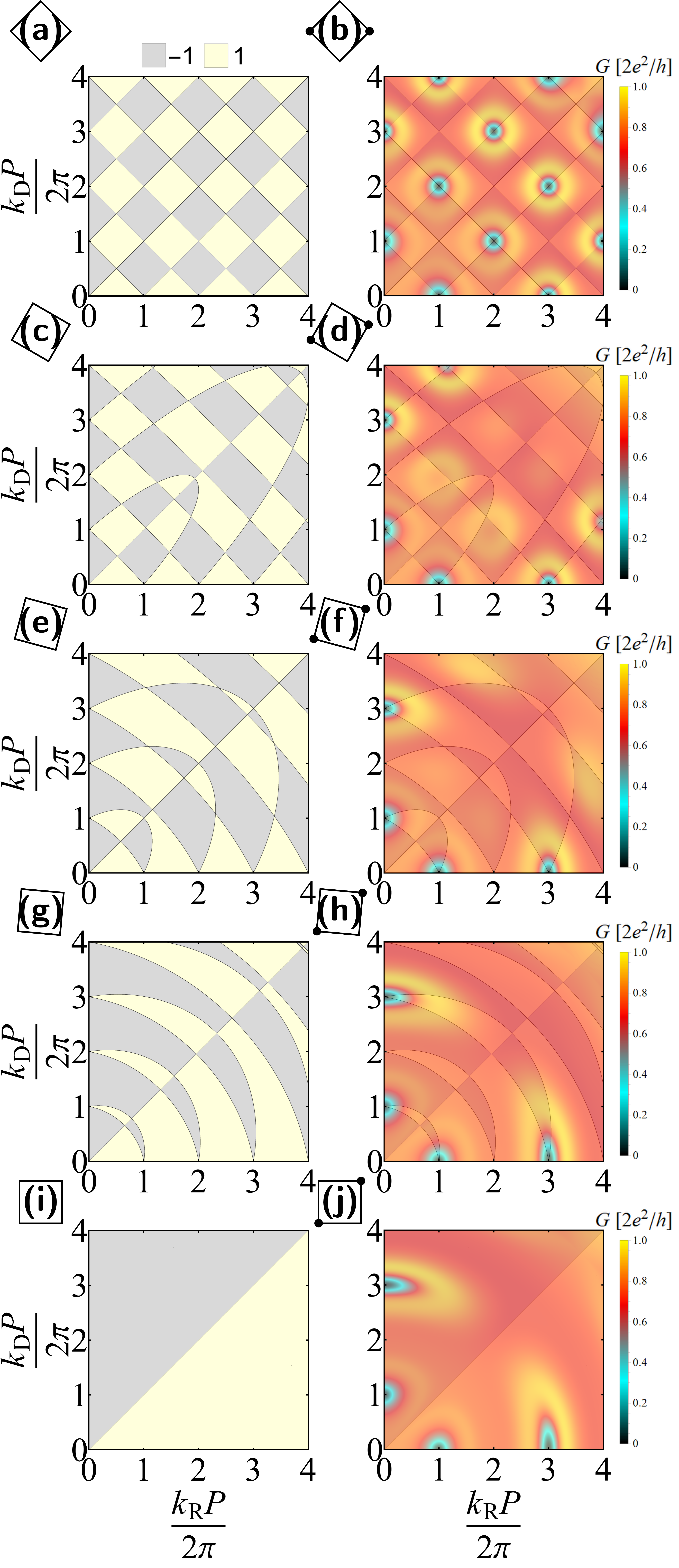}
    \caption{\label{Windingsquarerot} \textbf{Comparison of the winding number and conductance for the square geometries with different orientations: } Winding number $\omega$ (left column) and winding number overlapped with the conductance in units of $2e^2/h$ (right column) for a square in its original configuration (a-b), a square rotated an angle $\eta=\pi/12$ (c-d), $\eta=\pi/6$ (e-f), $\eta=2\pi/9$ (g-h), $\eta=\pi/4$ (i-j) subject to Rashba and Dresselhaus SOI. The left (right) dot in the sketches of the polygonal structures in panels (b), (d), (f), (h) and (j), indicates the position of the input (output) lead.}
\end{figure}
%
%

In Fig.~\ref{Windingsquarerot} we present the winding number and the conductance for the case of the square loop for additional values of the rotation angle $\eta$ with respect to the crystallographic axes. We can clearly observe the smooth transition from the checkerboard pattern in the first row to the \emph{featureless} structure in the last row as $\eta$ goes to $\pi/4$. When $\eta=0$, the field texture has  discontinuities with an angle $\pi/2$ that persist at the square's vertices for any field configuration (except when Rashba and Dresselhaus SOIs are equal and a persistent spin helix is set up). As $\eta$ increases, field discontinuities change their angle for different field configurations: optimal $\pi/2$ field-texture discontinuities at the vertices persist only along the Rashba and Dresselhaus axes in Figs.~\ref{Windingsquarerot}(a) to ~\ref{Windingsquarerot}(j), where a definite interference pattern in the conductance survives, as discussed in Refs.~\cite{Bercioux_2005A,Hijano_2021,Rodriguez_2021} for the Rashba case. 
Away from the Rashba and Dresselhaus axes in Figs.~\ref{Windingsquarerot}(a) to ~\ref{Windingsquarerot}(j), field-texture discontinuities are softened. Consequently, spin scattering is reduced, discouraging the development of complex spin textures with changing winding numbers. Moreover, counter-propagating spin carriers tend to gather equal phases, and destructive interference is suppressed.

\section{Case of other polygonal structures}
%
%
\begin{figure}[!ht]
    \includegraphics[width=0.8\columnwidth]{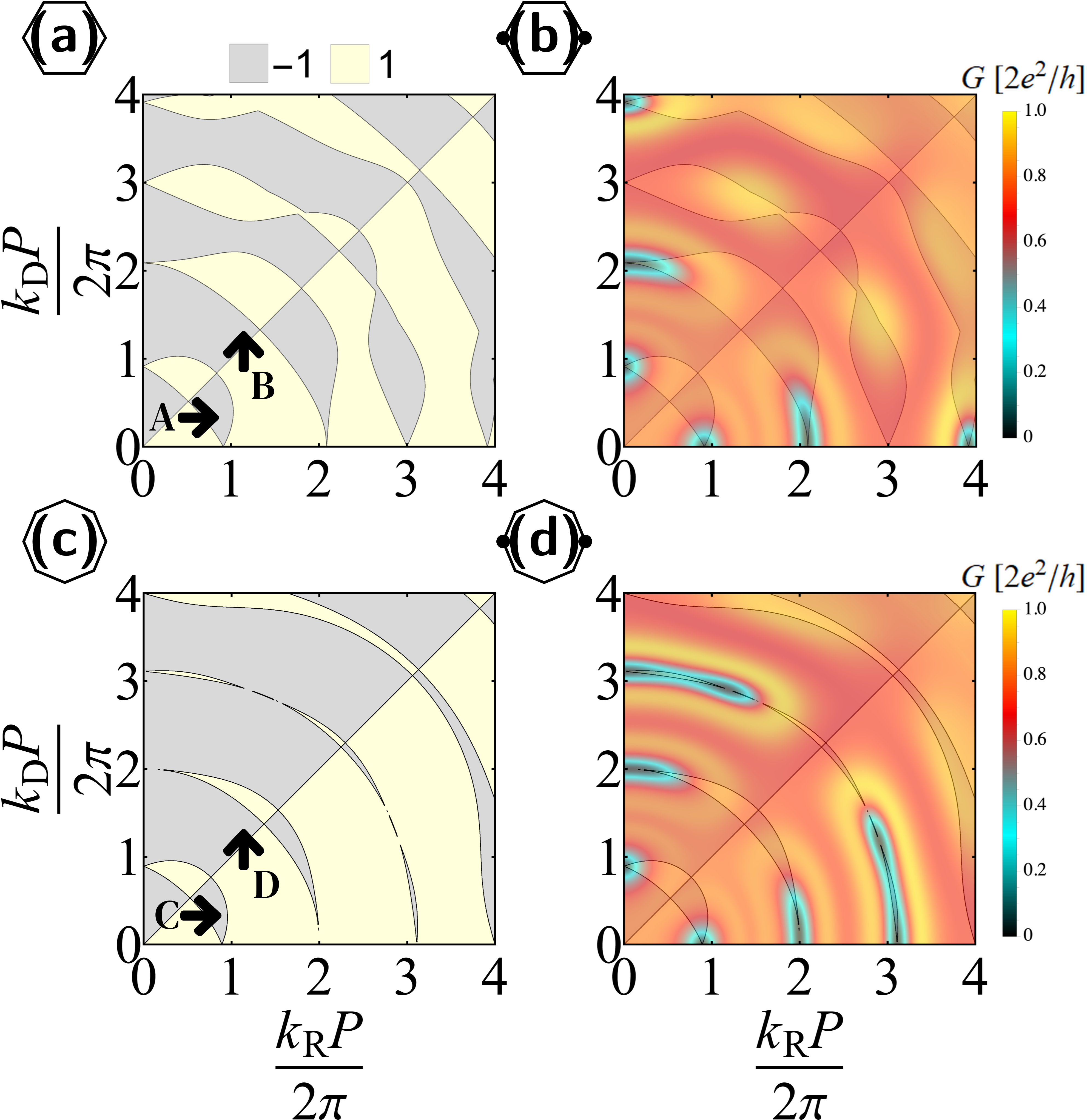}
    \caption{\label{Windingother}\textbf{Comparison of the winding number and conductance for the hexagon and the octagon geometries: } Winding number $\omega$ (left column) and winding number overlapped with the conductance in units of $2e^2/h$ (right column) for a hexagon (a-b) and an octagon (c-d) subject to Rashba and Dresselhaus SOI. The marks A and B in panel (a) and C and D in panel (c) refer to the position where we evaluate the winding number in Fig.~\ref{Texturesother}. The left (right) dot in the sketches of the polygonal structures in panels (b), and (d), indicates the position of the input (output) lead.}
\end{figure}
%
%
In Fig.~\ref{Windingother}, we show the conductance map of polygonal structures different from the one presented in the main text, specifically, for a hexagon and an octagon. 
As shown in the main text, the conductance is symmetric with respect to the critical line $k_\text{R}=k_\text{D}$. The hexagon and the octagon show a structure similar to the fishbone pattern shown by the ring.
The Rashba SOI and Dresselhaus SOI modify the SU(2) phase acquired by the electrons when traveling through these polygons; this leads to a shift in the position of the conductance maxima and minima with respect to the SO field strengths~\cite{Bercioux_2005B}. This quantum phase arises from the electron spin precession around the effective magnetic field associated with the SOIs; in the case of pure Rashba SOI, this effect is known as the Aharonov-Casher effect~\cite{AharonovCasher_1984}. The spin-phase gathering is controlled by two different scales: the perimeter and the edge lengths of the polygon. This is reflected in the oscillations of the conductance, which shows broader and narrower maxima for different values of $k_{\mathrm{R}}P$ and $k_{\mathrm{D}}P$ associated with two different frequencies~\cite{Bercioux_2005B,Hijano_2021}.
%
%
\begin{figure}[!h]
\centering
    \includegraphics[width=0.8\columnwidth]{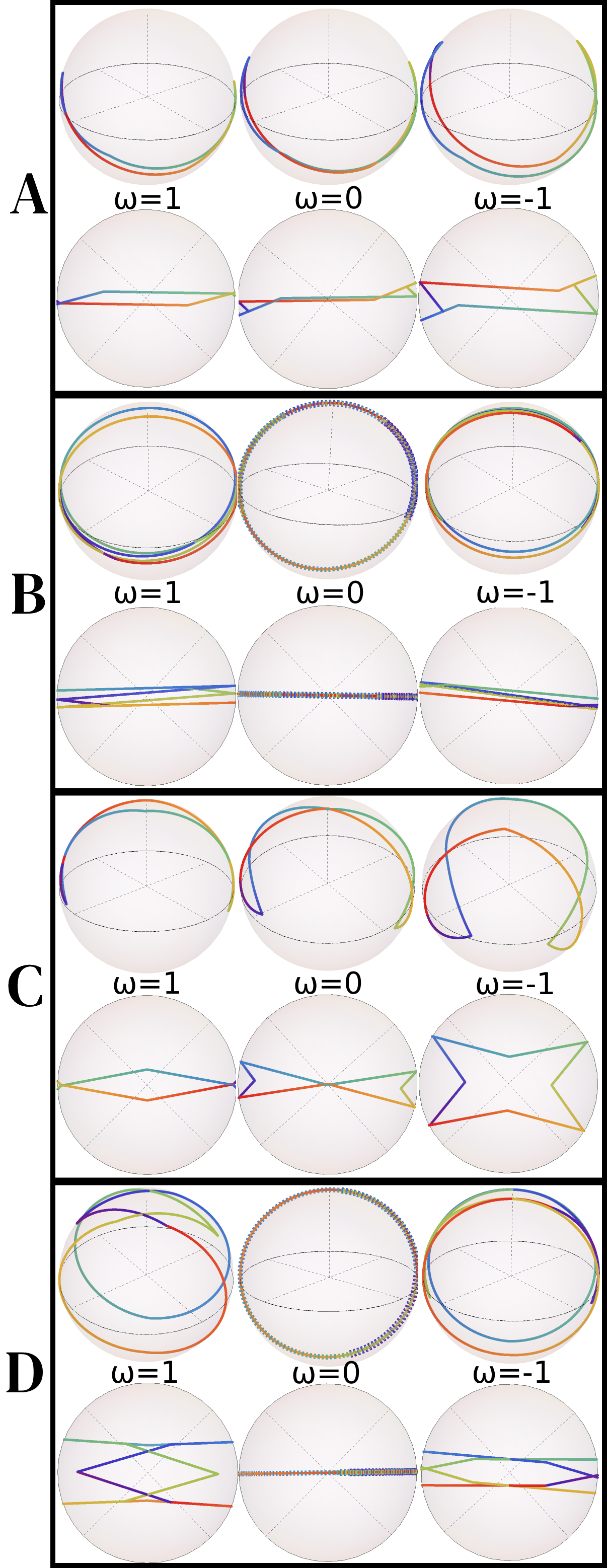}
    \caption{\label{Texturesother}\textbf{Spin texture for the hexagon and the octagon geometries: } The spin texture of a propagating mode in the Bloch sphere (up) and its azimuthal projection (down) for different spin-orbit interaction strengths. The spin-orbit interaction strengths are given by lines A and B (hexagon), C and D (octagon) in Fig.~\ref{Windingother}. In each panel, the colour indicates the circulation of the local spin states as the carrier propagates along the perimeter from red to violet.}
\end{figure}
%
%
In the pure Rashba SOI limit, the periodicity of the broader maxima is related to the edge lengths, where the period is $N\pi$. This period tends to infinity as the number of edges $N$ tends to infinity, so the broad maxima disappear for the ring, apart from the one located at the origin. The periodicity of the narrow maxima is related to the length of the perimeter; therefore, it has a weaker dependence on the number of edges of the polygon~\cite{Rodriguez_2021}. The period of the conductance oscillations ranges from $4\pi$ for the case of the square to $2\pi$ as the number of edges and the Rashba SOI strength increase. Oscillations of period $2\pi$ are identified with the adiabatic limit: when the dimensionless SOI strengths tend to infinity, we obtain the adiabatic limit in which the spin is aligned with the effective magnetic field during transport, and Berry phases arise~\cite{Bercioux_2005B,Frustaglia_2004,Rodriguez_2021}. Adiabatic spin transport is never really achieved in polygons, where vertices act as spin-scattering centers due to the abrupt change of direction of the SOI at the vertices of polygons.

%
%

%
%
The spin textures for hexagonal and octagonal loops are represented in Fig.~\ref{Texturesother}. For both polygonal loops, the textures mirror the results shown in the main text, where we obtain antisymmetrical textures with respect to the critical line. The textures tend to follow the dominant field and share its topological characteristics. The patterns of anomalous winding (explained in detail in the main text) are correlated with the conductance, and they become closer to the fishbone pattern of the ring as the number of sides of the polygon loop increases.
Figures~\ref{Texturesother}(A),~\ref{Texturesother}(B),~\ref{Texturesother}(C), and~\ref{Texturesother}(D) illustrate the winding transitions taking place in these textures. Similarly to the case of the ring loops and in contrast to the case of the square loops, a winding transition aside from the critical line does not require a full collapse of the spin texture with vanishing geometric phases. 

\end{document}